\documentclass[journal=jacsat,manuscript=article]{achemso}

\usepackage{graphicx}
\usepackage{mathtools}
\usepackage{amsmath}
\usepackage{epsf}
\usepackage{lipsum}
\usepackage{verbatim}
\usepackage{braket}
\usepackage[version=4]{mhchem}
\usepackage{bm}
\usepackage{bbold}
\usepackage{color}
\usepackage{xcolor}

\newcommand{\RN}[1]{%
 \textup{\uppercase\expandafter{\romannumeral#1}}%
}

\definecolor{mymagenta}{RGB}{200, 0, 100}
\definecolor{myblue}{RGB}{45, 48, 146}

\graphicspath{{figures/}}

\newcommand{\RNum}[1]{\uppercase\expandafter{\romannumeral #1\relax}}
\DeclareMathOperator*{\argmax}{arg\,max}

\newcommand{\be}{\begin{equation}}
\newcommand{\ee}{\end{equation}}
\newcommand{\ba}{\begin{eqnarray}}
\newcommand{\ea}{\end{eqnarray}}
\newcommand{\baa}{\begin{eqnarray}}
\newcommand{\eaa}{\end{eqnarray}}
\newcommand{\ed}{\end{document}}

\author{Hakon Volkmann}
\email{hakon.volkmann@physik.hu-berlin.de}
\author{Raamamurthy Sathyanarayanan}
\author{Alejandro Saenz}
\affiliation{\textit{AG Moderne Optik, Institut f\"{u}r Physik, Humboldt-Universit\"{a}t zu Berlin, Newtonstra{\ss}e 15, 12489 Berlin, Germany}
}
\author{Karl Jansen}
\affiliation{CQTA, DESY Zeuthen, Platanenallee 6, 15738 Zeuthen, Germany}
\alsoaffiliation{
 Computation-Based Science and Technology Research Center, The Cyprus Institute, 20 Kavafi Street,
2121 Nicosia, Cyprus
}
\author{Stefan Kühn}
\affiliation{CQTA, DESY Zeuthen, Platanenallee 6, 15738 Zeuthen, Germany}
\alsoaffiliation{
 Computation-Based Science and Technology Research Center, The Cyprus Institute, 20 Kavafi Street,
2121 Nicosia, Cyprus
}

\title{\color {black}A qubit-ADAPT Implementation for H$_{\bf 2}$ Molecules 
using an Explicitly Correlated Basis }

\abbreviations{IR,NMR,UV}

\begin{document}

\begin{abstract}
With the recent advances in the development of devices capable of performing
quantum computations, a growing interest in finding near-term applications
has emerged in many areas of science. In the era of non-fault tolerant
quantum devices, algorithms that
only require comparably short circuits accompanied by high repetition rates
are considered to be a promising approach for assisting
classical machines with finding solution on computationally hard problems.
The ADAPT approach previously introduced in Nat.\ Commun.\ {\bf 10},
3007 (2019) extends the class of variational quantum eigensolver
(VQE) algorithms with dynamically growing \emph{ans\"atze} in order to find approximations
to ground and excited state energies of molecules. In this work, the ADAPT
algorithm has been combined with a first-quantized formulation for the 
hydrogen molecule in the Born-Oppenheimer approximation, employing the 
explicitly correlated basis functions introduced 
in J.\ Chem.\ Phys.\ {\bf 43}, 2429 (1965). By the virtue of their explicit
electronic correlation properties, it is shown in classically performed simulations
that relatively short circuits yield chemical accuracy ($< 1.6$ mHa) for ground and excited state potential
curves that can compete with second quantized approaches such as Unitary Coupled Cluster. 
\end{abstract}

\section{Introduction}\label{sec-introduction}
Hybrid quantum-classical variational quantum algorithms (VQAs) have become a
major research subject in recent years~\cite{qca:cere21} that promise to provide near-term applications for
the growing number of available, non-fault tolerant gate-based
quantum computers, commonly called noisy intermediate-scale quantum (NISQ) devices~\cite{qca:pres18}. This kind of devices is succumbs to the limited
 number and the limited connectivity of qubits, decoherence times and gate fidelity. As such, most of the original quantum algorithms developed in the last decades still
cannot be realistically implemented with the hardware available today.
Hence, instead of implementing deep and complex circuits, the strategy considered most promising in designing algorithms 
suitable for such devices focuses on compact circuits at the price of an increased number of measurements. A classical computation back-end then
may be employed to ``guide'' the quantum device through the Hilbert-space and iteratively improve the obtained solution.
This idea of augmenting parts of a classical optimization algorithm by replacing them with small quantum circuits
on NISQ devices gained considerable interest~\cite{qca:arut20} and lead to numerous investigations of
potential applications in the areas of combinatorial optimization~\cite{qca:farh14}, high energy~\cite{qca:func21, qca:clem22}, solid state~\cite{qca:sher21, qca:feul22} and particularly 
chemical physics~\cite{qca:yung14,qca:peru14,qca:hemp18}, just to name a few. As the Rayleigh-Ritz principle is offering strict upper bounds
to ground-state energies of (semi-)bounded Hermitian operators, many quantum-chemistry problems can be 
formulated in terms of a variational parameter optimization.\\
The intricate interplay of electronic-vibronic correlation beyond the adiabatic Born-Oppenheimer description results in an
exponentially growing Hilbert-space, whose representation quickly becomes intractable on classical computers. This renders
the search for an accurate description of molecular structure a
particularly challenging task and sets a major drive for research on quantum algorithms in the context of
quantum chemistry. The variational quantum eigensolver (VQE) algorithm for example considers the energy cost-function $E(\bm\theta)$
of some Hamilton operator $\hat H$,
\be
  \label{eq:costfn}
  E(\theta) = \braket{\psi(\bm\theta)|\hat H|\psi(\bm\theta)},
\ee
such that a $\bm\theta$-parametric quantum circuit prepares the trial state $\ket{\psi(\bm\theta)}$ and performs projective
measurements in order to estimate $E(\bm\theta)$ statistically. This result can be used as a data point for
a classical non-linear optimization routine that seeks to predict improved values of $\bm\theta$ in an iterative manner. Another common approach
would apply the (variational) quantum imaginary-time evolution method~\cite{qca:mcar19} in order to systematically propagate
$\ket{\psi(\bm\theta)}$ to a reasonable ground-state approximation that will be used to estimate Eq.(\ref{eq:costfn}) once.
With the availability of well-tested and decades-long developed classical numerical methods such as 
Coupled Cluster, Configuration Interaction (CI), M{\o}ller-Plesset, or Kohn-Sham Density Functional Theory,
it remains an open question
whether such VQAs can give an actual quantum advantage compared to the former methods~\cite{qca:gont22, qca:bitt21, qca:elfv20}.
Yet, the accelerating speed of advances in the development of quantum hardware
shows that gathering experience with novel kinds of algorithms is becoming increasingly
relevant today and in the future.\\
The first experimental implementations~\cite{qca:yung14, qca:peru14} of VQE algorithms
on NISQ devices utilized a trotterized version of an Unitary Coupled-Cluster (UCC)
\cite{qca:taub06} ansatz
\be
 \label{eq:tucc}
 \ket{\psi(\bm\theta)} = \prod\limits_i \mathrm e^{\theta_i(\hat\tau_i-\hat\tau_i^\dagger)} \ket{\phi_0},
\ee
where $\bm\theta$ denotes the vector of the (real) variational parameters. The product runs over 
all possible single, double, etc. excitations from a Hartree-Fock reference determinant encoded by $i$, while
$\hat\tau_i$ describes the corresponding fermionic annihilation operator 
and $\ket{\phi_0}$ is a reference state. As there is no efficient classical algorithm known for
calculating Eq.(\ref{eq:tucc}) without introducing further approximations~\cite{qca:peru14, qca:rome18},
the ansatz state will be prepared and its energy in Eq.(\ref{eq:costfn}) will be measured on a NISQ device instead.\\
In principle, the ansatz can be chosen arbitrarily, 
provided it can be systematically converged to the true state by
increasing the number of variational parameters. While chemically-motivated ans\"atze as in Eq.(\ref{eq:tucc})
likely cover most of the relevant parts of the Hilbert-space with a simple parametrization, their implementation on a quantum
device usually involves the conversion of the the excitation operators $\hat\tau_i$ from
the fermionic Fock space to the spin space by using e.g. Jordan-Wigner~\cite{qca:jord28} or 
Bravyi-Kitaev~\cite{qca:brav02} transformations. The fermionic exchange symmetry, expressed in the language of
Pauli-operators $\hat\sigma_q, q\in\{0, x, y, z\}$, results in lengthy Pauli-operator products or ``strings'',
that quickly can grow intractable for a NISQ device. While alternative, more hardware-efficient
ans\"atze (HEA)~\cite{qca:kand17} are possible, their disadvantage lies in the necessity to explicitly
enforce proper (anti)symmetry and particle-conservation properties via artificial penalty terms~\cite{qca:mccl16},
which in practice will
have to be re-acquired by the ansatz, leading to an increased number of variational parameters~\cite{qca:till22}. 
Additionally, it is argued that highly expressible ans\"atze~\cite{qca:sim19} such as the HEA cause 
\emph{Barren plateaus}~\cite{qca:mccl18, qca:cere21a, qca:tuys22} in the cost-function: the effect of an
exponentially vanishing gradient and its resulting ``narrow gorge'' in the cost-function landscape
considerably increases the shot count in order to retain the required precision for navigation.\\
The adaptive derivative-assembled pseudo-Trotter / problem-tailored (ADAPT) approach originally proposed by Grimsley et al.\cite{qca:grim19}
is an iterative extension to the VQE algorithm outlined above. By featuring a repeated alteration of the ansatz and,
correspondingly, its parameter landscape
at each iteration, it acquires an intrinsic
resilience to Barren plateaus~\cite{qca:grim23}. The combination of shallow
circuits and systematic convergence properties renders this a particularly
promising candidate for future NISQ applications and thus has been extensively
studied in recent works~\cite{qca:tang21, qca:arma22, qca:rome22}. However, to the knowledge of the authors, 
so far convergence (and by extension accuracy) of ADAPT only has been established in terms of minimal basis sets using Slater type orbitals
approximated by a linear combinations of three Gaussian functions (STO-3G). In order to study
effects of electronic correlation, such basis sets are known to be insufficient and only can serve as a rough estimate usually
mostly at the Hartree-Fock level. In contrast, explicitly correlated (EC) basis functions~\cite{qca:kong11} provide highly
accurate energies and give a more realistic estimate on the efficiency of the method. Despite of their excellent convergence
behavior, EC methods in VQAs so far have only attracted minor attention~\cite{qca:schl22}. Thus the goal of this work is to
investigate the combination of ADAPT with EC methods for solving the electronic-structure problem for bound states of molecular
hydrogen in first quantization. Although the electronic properties of H$_2$ are well-understood~\cite{qca:shar70},
the diatomic molecule still serves as a reasonable benchmark for estimating quantum resources required to reach 
chemical accuracy~\cite{qca:pete12} for molecules of
greater complexity and less symmetry. The set of two-electron functions
proposed by Ko{\l}os et al.~\cite{qca:kolo65} is unrivaled in terms of accuracy for molecular hydrogen.
Hence, it will constitute the basis set for the computation of ground and excited state potential
energy curves using the variational quantum deflation (VQD) technique~\cite{qca:higg19}, which has been successfully applied using
ADAPT in the past~\cite{qca:chan21,qca:yord22}.\\
The paper is organized as follows.
In sec. II, ADAPT is introduced and its usage for first quantized systems being motivated, whose implementation details 
with first-quantized two-electron orbitals will be elaborated in Sec. III. In Sec. IV A,
resulting data from the potential curves of the Hydrogen molecules using 4, 5 and 6 qubits will be presented. Finally, in Sec. V the results will be discussed.

\section{ADAPT-VQE}
The ADAPT approach features a compact dynamical ansatz that comes with an extension
of the VQE procedure outlined above. At the $M$th iteration, the trial state
\be
  \ket{\psi_M}=\prod\limits_{j=1}^M\mathrm e^{\mathrm i\theta_j \hat P_j}\ket{\psi_0},\quad \hat P_j = \bigotimes\limits_{k=1}^n \hat\sigma^{(k)}_{q(j)},
\ee
consists of a product of $M$ unitary operations with corresponding parameters $\theta_j\in\mathbb R$ that serve as
variational parameters on an initial reference state $\ket\psi_0$. The generators $\hat P_j$ are chosen
from a non-exhaustive ``pool'' of pre-determined Pauli strings, where $\hat\sigma^{(k)}_{q(j)}$ is the $k$th-qubit Pauli-operator for $q(j)\in\{0,x,y,z\}$
in the $n$-qubit Pauli string $\hat P_j$.\\
At the $M+1$th iteration of the algorithm, the ansatz
is extended by an operator chosen from the pool 
such that the most relevant part of the Hilbert-space in terms of
energy-gradient impact is made available 
where the dynamically-sized vector $\bm\theta=(\theta_1,\ldots,\theta_{M+1})^{\rm T}$ is optimized on. This can be done by
selecting the particular $\hat P_j$ out of the pool of $2n-2$ operators such that
\begin{align}
 \label{eq:select}
 j=\argmax\limits_{k\in\{1,\ldots,2n-2\}}\Bigg|\frac{\partial}{\partial\theta}\langle\psi_M|\mathrm e^{-\mathrm i\theta\hat P_k^\dagger} \hat H \mathrm e^{\mathrm i\theta\hat P_k}| \psi_M\rangle\Big|_{\theta=0}\Bigg|=\argmax\limits_{k\in\{1,\ldots,2n-2\}}\Big|\langle\psi_M| [\hat H,\hat P_k]|\psi_M\rangle\Big|.
\end{align}
Hence, not only the energy but also its gradient w.r.t. all pool operators will be measured on the quantum device.
Once an operator has been selected from the pool, the newly obtained ansatz will be subject to
a VQE optimization procedure that yields an energy and its corresponding parameter vector $\bm\theta$.
This procedure is repeated until the norm of the energy-gradient vector falls below a defined
threshold, upon which the iteration stops. For a more detailed introduction to the ADAPT method the reader may be referred to the work of Grimsley \emph{et al.}~\cite{qca:grim19}.\\
The operator pool was originally composed of single and double excitation operators $\hat\tau_i$ which 
got mapped to the qubit space by the Jordan-Wigner (JW) transformation. This way, it has been demonstrated that
adaptive ans\"atze in fact outperform ``traditional'' approaches such as the k-fold products of
unitaries paired with generalized single and double excitations (k-UpCCGSD) \cite{qca:lee18} in terms of both gate counts
and errors.\\
It was later shown~\cite{qca:tang21} that a more hardware-efficient variant of the ansatz, coined qubit-ADAPT,
is able to achieve comparable convergence as the originally proposed ``fermionic'' ADAPT. Instead of just including such Pauli strings
that result from JW-transformed fermionic excitation operators, the class of admissible Pauli strings is extended to all which are able to
span the dynamical Lie algebra of rotations in the qubit-space. Despite of featuring longer
ans\"atze and, correspondingly, more ADAPT iterations, the overall circuit depth is still largely reduced,
similarly to the HEA, but without the issues of (expressibility-induced)
Barren plateaus~\cite{qca:grim23}. Interestingly, it has been shown that the size of the operator pools dealing as the source of possible ansatz extensions 
can be as small as $2n-2$~\cite{qca:tang21, qca:shko21}, with $n$ being the number of qubits while still maintaining reachability of all real states
in the $n$-qubit Hilbert-space, which drastically reduces the number of
circuits and gradient measurements required for optimization.\\
As already mentioned, previous work on VQE algorithms for chemical problems focused mostly on formulations described
in second quantization.
While it has been argued that first quantization approaches are in fact inferior for treating electronic-structure problems
on quantum computers~\cite{qca:till22}, they come with the property of fermionic exchange antisymmetry being
contained inside the states and the matrix elements rather than in the operators. Correspondingly, a qubit-ADAPT
ansatz expressed in terms of first quantization
would neither have to deal with particle number or exchange symmetry explicitly, nor would it be
needed to re-acquire such properties by an increased number of variational parameters that have to be trained, as no fermion-to-spin transformations
have to be performed. This motivates to implement a first quantized variant of the qubit-ADAPT approach. The following section will illustrate how to address the H$_2$ molecule in such a manner.

 \section{Implementation}\label{sec-kolos}
 In order to efficiently account for the correlation effects in molecular hydrogen,
 explicitly correlated basis functions expressed in prolate spheroidal
 coordinates~\cite{qca:vann04} $\mathbf r_i=(\xi_i,\eta_i,\phi_i)$ are employed.
 The Ko{\l}os-Wolniewicz basis~\cite{qca:kolo65} expresses the adiabatic wave function
 $\psi(\mathbf r_1, \mathbf r_2;R)$, parametrized by fixed values of the internuclear distance $R$,
 as 
 \begin{align}
  \begin{split}
   \psi(\mathbf r_1, \mathbf r_2;R) &= \sum\limits_i c_i(\Phi_i(\mathbf r_1,\mathbf r_2;R)x_1^l\pm\Phi_i(\mathbf r_2,\mathbf r_1;R)x_2^l),
  \end{split}
 \end{align}
 where $c_i$ denotes an expansion coefficient, $x_k$ is the $k$th electron Cartesian coordinate perpendicular
 to the molecular axis, $l=0,1,\ldots$ determines $\Sigma,\Pi,\ldots$ inter-molecular-axis projected
 angular momentum symmetries
 and the $\pm$ sign determines the spin multiplicity for singlet and triplet states, respectively.
 The functions $\Phi_i(\mathbf r_1,\mathbf r_2;R)$ are defined as
 \begin{align}
     \Phi_i(\mathbf r_1, \mathbf r_2;R) = \sum\limits_i c_i \mathrm e^{-\alpha\xi_1+\overline{\alpha}\xi_2}
   \cosh(\beta\eta_1+\overline{\beta}\eta_2)\xi_1^{r_i}\eta_1^{s_i}\xi_2^{\overline{r}_i}\eta_2^{\overline{s}_i}\left(\frac{2r_{12}}{R}\right)^{\mu_i},
 \end{align}
 where $r_{12}$ denotes the inter-electronic distance, $r_i,\overline{r}_i,s_i,\overline{s}_i,\mu_i\in\mathbb N$
 determine the order of the polynomial and the four real nonlinear parameters $\{\alpha,\overline{\alpha},
 \beta,\overline{\beta}\}$ can be used to further tune the energy
 for particular values of $R$ or a certain excited state.\\
 Written this way,
 the basis set is not orthogonal with respect to the $L^2$ scalar product. Hence, the basis is further
 Gram-Schmidt orthonormalized in order to both save oneself from explicitly treating
 the overlap matrix $\braket{\Phi_i|\Phi_j}$ and to remedy numerical near-linear dependency problems
 that potentially arise for sufficiently large bases at a limited numerical precision~\cite{qca:kolo75}.
 For a fixed nonlinear quintuplet
 and maximal polynomial order, the Hamiltonian matrix elements $h_{ij}=\braket{\Phi_i|\hat H|\Phi_j}$ 
 are computed classically for all values of internuclear distance $R$ using
 an efficient recursive integration scheme~\cite{qca:kolo65, qca:rued56}.\\
 
 \noindent The qubit-representation of the Hamilton operator is obtained
 by simply employing a binary encoding representation~\cite{qca:till22}
 \begin{align}
  \label{eq:hambin}
  \hat H=\sum\limits_{i,j=1}^N h_{ij}\ket{{\rm bin}(i)}\bra{{\rm bin}(j)},
 \end{align}
 where $\ket{{\rm bin}(i)}\bra{{\rm bin}(j)}\simeq\ket{\Phi_i}\bra{\Phi_j}$ is represented by Pauli strings
 of length in $O(\log_2(N))$. Due to the first quantized approach, neither
 the fermionic exchange (or in fact any molecular) symmetry, nor a distinction between one- and 
 two-particle states have to be explicitly incorporated, reducing the circuit
 depth for obtaining expectation values of $\hat H$ by, \emph{e.g.}, statistical sampling from 
 repeated measurements of the individual terms.\\
 At each iteration, the pool operators $\hat P_j\in\mathcal P$ are chosen from a
 complete minimal pool~\cite{qca:shko21} of size $2n-2$ whose definition can be found in the supplementary material.
 The question which selection criterion should be applied for picking the pool operators has
 been previously discussed, which includes considerations of utilizing maximum gradient norm~\cite{qca:grim19, qca:grim23, qca:bert22} or
 maximum impact on energy~\cite{qca:arma22}. While the latter approach was reported~\cite{qca:arma22} to show superior convergence properties, this work
 solely utilized the former one as in Eq.(\ref{eq:select}) for the sake of comparability with the majority of previous works.
 The $k$th component of the energy-gradient $\nabla E$ can be rewritten as~\cite{qca:grim19}
 \begin{align}
 \label{eq:grad}
 (\nabla E)_k=\frac{\partial}{\partial\theta}\langle\psi_M|\mathrm e^{-\mathrm i\theta\hat P_k^\dagger} \hat H \mathrm e^{\mathrm i\theta\hat P_k}| \psi_M\rangle\Big|_{\theta=0} =
 2\ \mathrm{Re}\ \{\bra{\psi_M}\hat H\hat P_k\ket{\psi_M}\}.
 \end{align}
 Hence, the energy-gradient, supplementary to the energy, is measured on the quantum device and can be implemented using
 either a Hadamard test or more efficient circuits~\cite{qca:mita19}. Once the (Euclidean) gradient norm falls below a selected
 threshold at a certain iteration, the algorithm stops and the corresponding energy is returned,
 as laid out in the original description~\cite{qca:grim19}.\\
 Additionally to the ground-state, excited state energies are obtained via
 the VQD technique. In order to obtain the $g$th excited state $\ket{\psi_M^g}$,
 additional $g-1$ deflation terms
 \begin{align*}
  \hat H_{\rm VQD} = \hat H + 2\ \mathrm{Re}\left\{\sum\limits_{i=1}^{g-1}\beta_i \bra{\psi^g_M}\hat V_k\ket{\Phi_i}\braket{\Phi_i|\psi^g_M}\right\}
 \end{align*}
 will be added to the Hamiltonian in the gradient Eq.(\ref{eq:grad}), provided the lower excited states $\ket{\Phi_i}$ with $i<g$ are known and $\beta_i>0$ is chosen to be larger
 than the largest spectral gap of $\hat H$. These terms can be implemented and measured 
 on the quantum device, \emph{e.g.}, using a SWAP test~\cite{qca:buhr01}.\\
 Since gradient information on the energy-landscape has to be explicitly provided, the gradient-based
 Broyden–Fletcher–Goldfarb–Shanno (BFGS) algorithm of the SciPy package~\cite{qca:jone01}
 has been adopted for optimization of $\bm\theta$ at each iteration of the ADAPT procedure such that the previously found
 parameters are re-used in the successive iteration of ADAPT (``warm-start''~\cite{qca:grim23}). 
 The matrix elements $h_{ij}$ are extracted from a modified Fortran code that has been extensively tested
 in the past~\cite{qca:jons01,qca:froe87,qca:froe93}. The ADAPT algorithm is entirely implemented in Python using the PennyLane framework~\cite{qca:berg18}. In order
 to increase the computational efficiency, the analytic gradient approach laid out in the supplementary information of the initial proposal~\cite{qca:grim19} has
 been implemented. 
 
 \section{Results}
 
 \begin{figure}
  \includegraphics[width=0.5\textwidth]{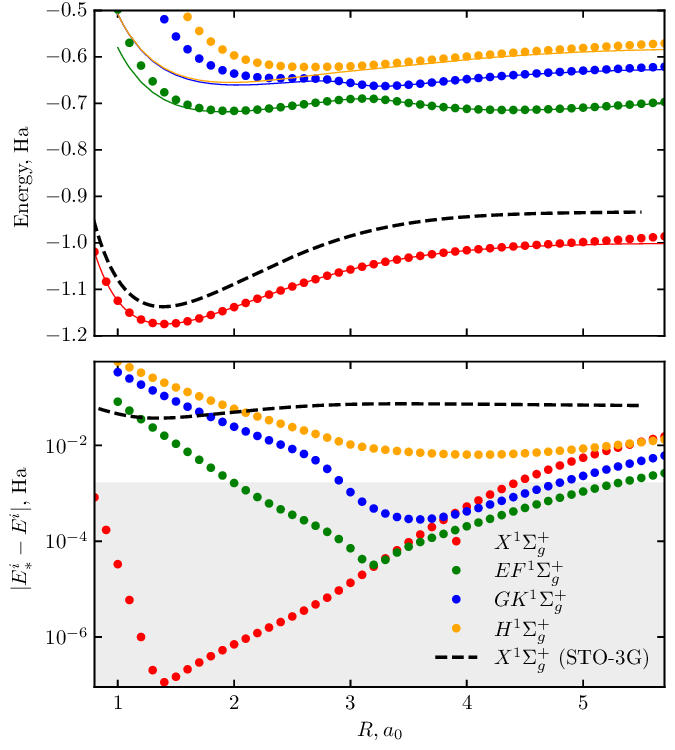}
  \caption{Top: Potential curves as obtained from qubit-ADAPT (dots) and highly accurate
  benchmark computations~\cite{qca:shar70} (solid lines) for the ground and first three excited states in the $^1\Sigma_g^+$ symmetry
  of H$_2$, indicated
  by the colors laid out in the legend. Bottom:
  Absolute difference between both results. The shaded area indicates chemical accuracy
  $<1.6$ mHa. The black (dashed) lines indicate the results obtained from a classical full-CI computation using a minimal Gaussian basis set (STO-3G). A basis size of 7 qubits has been used.}
  \label{fig:chac}
 \end{figure}
 
\begin{figure}
  \includegraphics[width=0.45\textwidth]{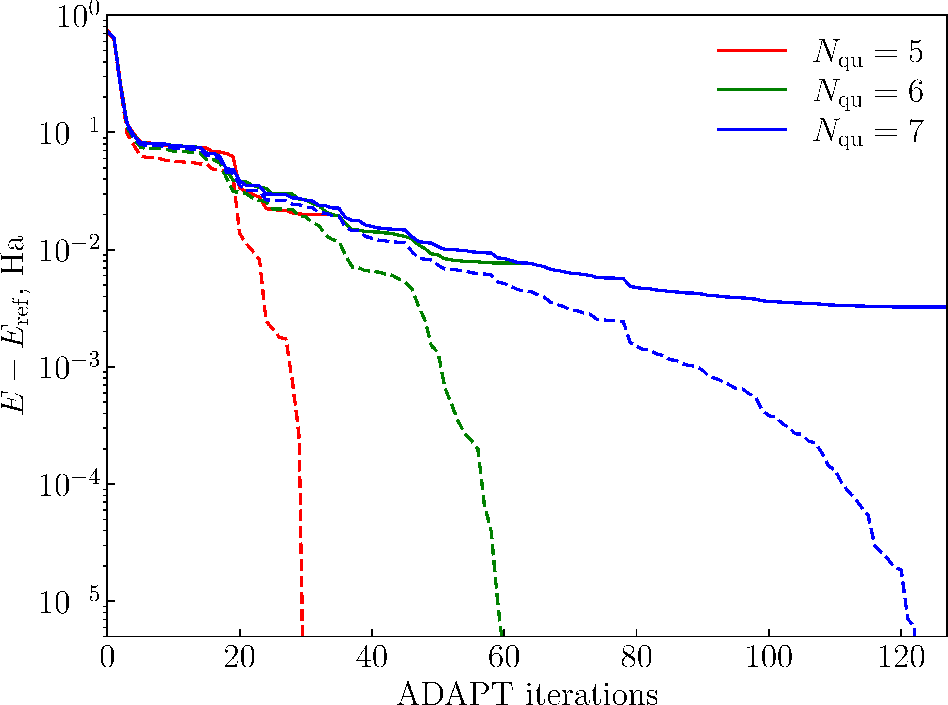}
  \caption{Comparison of ADAPT's $GK$-state energy convergence against two different benchmark energies $E_{\rm ref}$:
  literature benchmark value~\cite{qca:woln77} (solid lines) 
  and exact diagonalization (dashed lines) for $N_{\rm qu} = 5, 6$ and $7$ qubits (colors) at $R = 2.8\ a_0$. }
  \label{fig:bench}
 \end{figure}

 \noindent Computations have been performed with basis sizes ranging from 5 to 7 qubits.
 Fig.\ref{fig:chac} depicts the potential-energy curves for the first four $^1\Sigma_g^+$ symmetry states obtained
 from a 128 basis-function (7 qubit) ADAPT computation and compares 
 them to (chemically) accurate literature values~\cite{qca:shar70}. It should be emphasized that, in the absence of a noise model, the results only
 display the ideal limiting case in terms of accuracy. The non-linear parameters $\{\alpha,\overline{\alpha},\beta,\overline{\beta},\mu\}$ have been optimized for an accurate ground
 state ($X^1\Sigma^+_g$) energy at equilibrium distance ($R = 1.4 a_0$) and subsequently kept fixed, which also can be recognized 
 by the strong minimum at equilibrium distance for the absolute energy error of the $X$ state. Despite of this limitation,
 a wide range of the ground-state potential curves and significant parts of the ones of the first two excited state lie below the threshold of chemical accuracy
 and are even able to reproduce the double-well shape of the EF and GK states caused by avoided crossings~\cite{qca:shar70} due to the "change" between ionic and covalent character.
 This changes beginning with the H state, where despite of faithful convergence of the ADAPT iteration to the eigenvalues of the
 truncated Hilbert-space representation of $\hat H$, no parts of the potential curve are found to be within chemical accuracy. In order to underline the accuracy offered by the EC basis, the ground-state
 potential curve using a minimal Gaussian basis as obtained from a full-CI classical computation using the PySCF framework~\cite{qca:sun17} is also provided. It can be clearly seen that this basis set is not able to produce chemically accurate energies at any point of the potential curve. In light of this circumstance, it is important to distinguish the notion of chemical accuracy in this work from that in previous works~\cite{qca:grim19,qca:sapo22}, where such full-CI results served as a benchmark.\\
 \begin{figure}
  \includegraphics[width=0.5\textwidth]{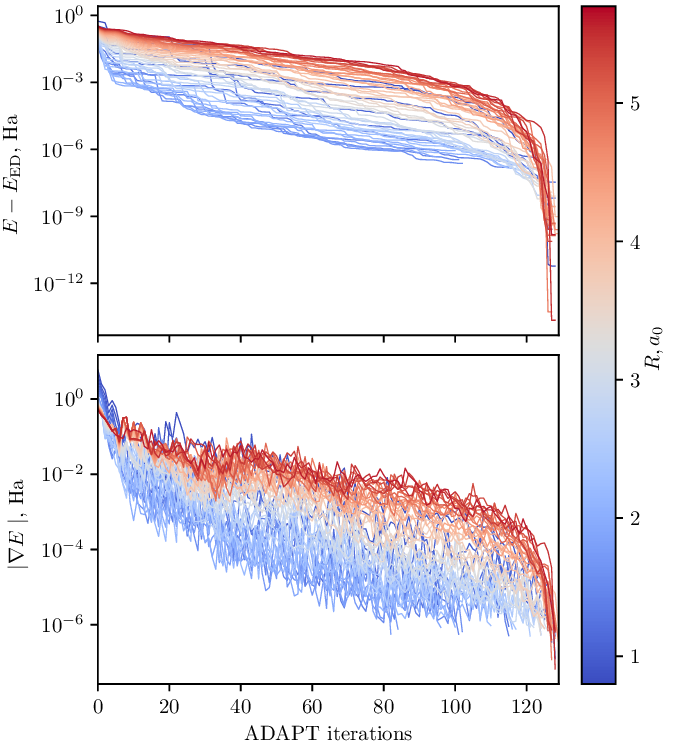}
  \caption{Top: Absolute difference between the energy obtained from the qubit-ADAPT VQE
  and exact diagonalization (ED) for the ground-state at each ADAPT iteration step.
  The different internuclear separations $R$ are indicated by colors ranging
  from blue for small to red for large values of $R$. Bottom: Euclidean norm
  of the adapt gradient Eq.(\ref{eq:grad}). A basis size of 7 qubits
  has been used.}
  \label{fig:conv_x}
 \end{figure}
 \begin{figure}
  \includegraphics[width=0.5\textwidth]{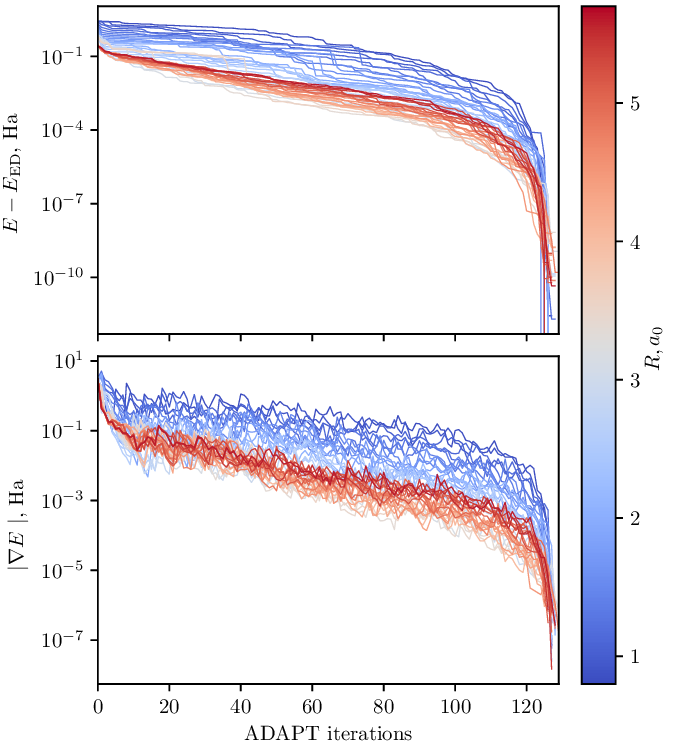}
  \caption{Same as Fig. \ref{fig:conv_gk}, but for the GK-state.}
  \label{fig:conv_gk}
 \end{figure}
 \noindent The convergence behavior of the $X$ and $GK$ states depending on the internuclear separation $R$ with respect to the ADAPT
 iteration count has been plotted in Figures
 \ref{fig:conv_x} and \ref{fig:conv_gk}, respectively. A similar convergence behavior for both absolute energy
 difference and gradient norm~\cite{qca:grim19,qca:tang21} can be recognized, where plateaus and and steps in the
 energy difference can be associated with areas of small and increasing gradient norms before a final drop in error
 indicates convergence. It should be noted that the benchmark energy is taken to be that from
 considering the truncated Hilbert-space approximation and not a true benchmark as in Fig.
 \ref{fig:chac}, as also being done in Refs.~\cite{qca:grim19,qca:shko21,qca:tang21}. Hence, it should be
 reiterated that convergence in this sense merely assures convergence in the sub-Hilbert
 space spanned by the qubit description, and only gives a necessary, but no sufficient condition
 on the viability of the method.
 \begin{figure*}
  \includegraphics[width=1.0\textwidth]{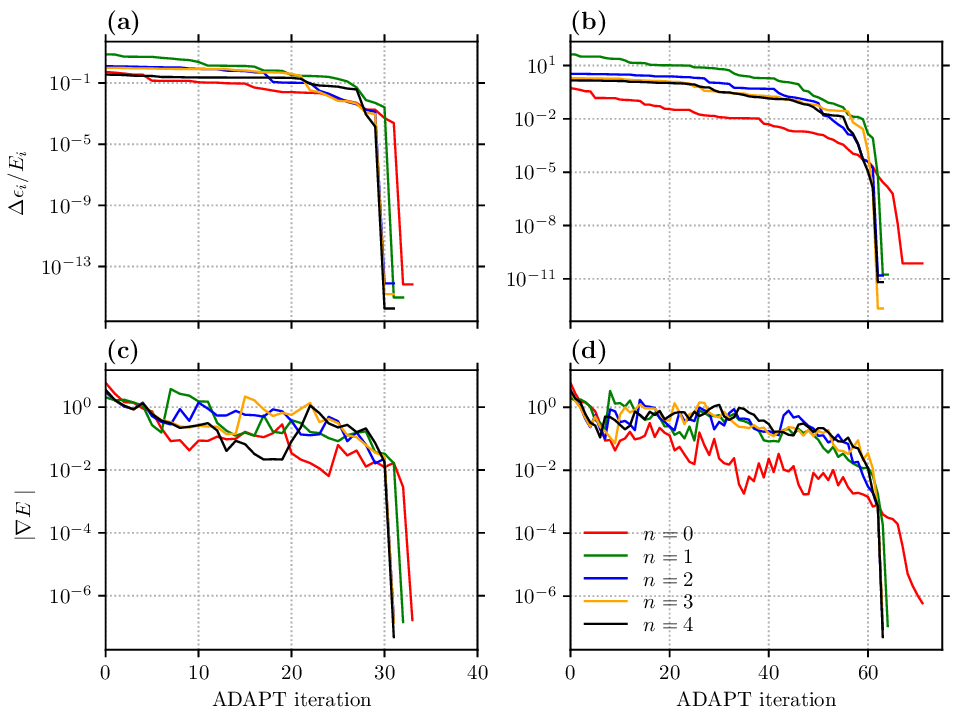}
  \caption{\label{fig:foo} Qubit-ADAPT relative energy errors (a), (b) and ADAPT gradient norm (c), (d)
  for 5 qubits (left) and 6 qubits (right) at internuclear distance $R = 0.8$ a.u.
  The individual excited states are identified by color as indicated in the legend and
  belong to the same $^1\Sigma_g^+$ symmetry. The ADAPT gradient threshold for convergence
  was selected at $10^{-6}$ a.u.; for the BFGS optimization tolerance a value of $10^{-7}$ has been used.}
 \end{figure*}
 \noindent Comparing the $R$-dependent convergence behavior between the two states, they all share the property of
 simultaneously reaching convergence at an iteration number which roughly seems to be close to the 
 truncated Hilbert-space dimension $2^n$. This is confirmed for 5 and 6 qubits in Fig. \ref{fig:foo}
 as well, where it is shown that both error and gradient norm collapse at around 32 and 64 iterations, respectively. However, while the overall trend of the gradient magnitudes $|\nabla E|$ tends to be
 anti-correlated with the internuclear separation $R$ for the $GK$ state in Fig. \ref{fig:conv_gk},
 the opposite behavior can be observed for the ground-state ($X$) in Fig. \ref{fig:conv_x}.
 This is also visible from the energy differences. As the initial reference state shares the same molecular
 symmetry just as all states in the considered Hilbert-space, none of the pool operators can mix states
 into the ansatz that would break the symmetry in favor of an improved energy estimate. Consequently,
 no emergence of ``gradient troughs'' as it has been observed in Refs.~\cite{qca:shko21,qca:bert22} can be seen, which is a beneficial feature of the first-quantized approach.\\
 Finally, Fig. \ref{fig:bench} compares the convergence behavior of the energies obtained from ADAPT to 
 two different energy-limit benchmarks for different basis sizes. One benchmark relative to the energy obtained
 from exact diagonalization (ED) in the truncated Hilbert-space accessible to ADAPT; the other being a reference literature value~\cite{qca:woln77}.
 Convergence with respect to the former hence only gives a statement on the consistency of the method,
 while only the latter allows to assess the overall performance. It can be seen that, initially, at low iteration counts,
 all curves are rather overlapping and show the same
 speed of convergence. While the energies converge to their ED limit at an iteration number corresponding to the Hilbert
 space dimension, it is clearly visible that the error with respect to the literature value is only slightly
 improved for the last iteration steps.
 
 \section{Discussion}
 \noindent
 As the minimal complete pools ensure that each real state in the approximating Hilbert
 space is able to be reached with a finite set of parameters~\cite{qca:tang21}, virtually no accuracy in
 the energies is lost when transitioning from an exact-diagonalization treatment on a classical device
 to a hybrid classical-quantum approach. This point usually is reached in an abrupt manner, manifest
 by the sudden drop in absolute error over multiple orders of magnitude at a certain ADAPT iteration number, as it
 has been frequently demonstrated~\cite{qca:grim19,qca:tang21,qca:grim23,qca:chan21} for different ADAPT implementations.
 As shown above, care must be taken when trying to deduce chemical accuracy from energy difference considerations.
 Even if the ADAPT iteration faithfully converges to its ED limit, this does not necessarily imply a particularly high accuracy 
 in energy in practice. As most of the work concerning ADAPT is employing minimal Gaussian basis sets such as STO-3G~\cite{qca:grim19,
 qca:tang21, qca:grim23, qca:anas22}, this
 fact should be kept in mind when drawing further conclusions~\cite{qca:sapo22}.
 Concomitantly, the gradient criterion employed for asserting convergence does not permit a definite statement about the
 accuracy of the obtained energy but rather a statement about the fidelity of the ansatz state on the approximated Hilbert
 space. Hence, the hydrogen molecule with its accurately characterized EC basis energies may provide a useful testing area for assessing 
 the actual performance of the method.\\
 In lieu of the fact that the number of parameters required for convergence always appears to grow with the size of
 the Hilbert-space dimension, it will become a significant question whether
 such a deep ansatz is actually necessary for obtaining chemical accuracy.
 Even in the case of a simple hydrogen molecule considered here this is a relevant issue, as the ADAPT ansatz is reminiscent of a generalized
 axis-angle parametrization of the target state, which requires $O(2^n)$ angles~\cite{qca:hoff72}. It can therefore be concluded
 that the success of applying ADAPT to molecules crucially depends on the ability of the algorithm to reach a
 satisfactory level of accuracy well before the point of convergence to the ED limit. That this is indeed the case
 for the highly efficient Ko{\l}os-Wolniewicz basis gives reason to believe that this can be achieved in practice.

\section*{Summary and Outlook}
In this work, a new approach of combining a first-quantized description of VQE employing EC basis functions with the qubit-ADAPT
algorithm has been presented. This way, the benefits of qubit-ADAPT VQE featuring small and trainable ans\"atze are combined with  
a straight-forward realization without the need of explicit anti-symmetrization. It was shown that qubit-ADAPT is capable of
fully exploiting the high efficiency of EC basis functions, as results for both ground and excited-state potential curves 
at chemical accuracy for large ranges of internuclear separation are demonstrated, which is beyond of what commonly used
minimal STO-3G basis sets can achieve. This provides a lower bound for the ansatz and problem
sizes required for describing simple diatomic molecules, which may be extrapolated to more complicated molecules featuring both
more electrons and reduced molecular symmetry. Such a description could be realized with more general EC methods such as
Gaussian geminals or F12 methods~\cite{qca:grot00}. Furthermore, the effects of quantum noise and its impact on convergence remains an open question
that only got little attention in the past and will be subject of future investigations.

\begin{acknowledgement}
We gratefully acknowledge financial support
from the Berlin Quantum Alliance and the Einstein Foundation (Perspectives of a quantum digital
transformation: Near-term quantum computational devices and quantum
processors, A: Project P6.
S.K.\ acknowledges financial support from the Cyprus Research and Innovation Foundation under project ``Quantum Computing for Lattice Gauge Theories (QC4LGT)'', contract no.\ EXCELLENCE/0421/0019.
This work is funded by the European Union’s Horizon Europe Framework Programme (HORIZON) under the ERA Chair scheme with grant agreement no.\ 101087126.
This work is supported with funds from the Ministry of Science, Research and Culture of the State of Brandenburg within the Centre for Quantum Technologies and Applications (CQTA). 
\begin{center}
    \includegraphics[width = 0.08\textwidth]{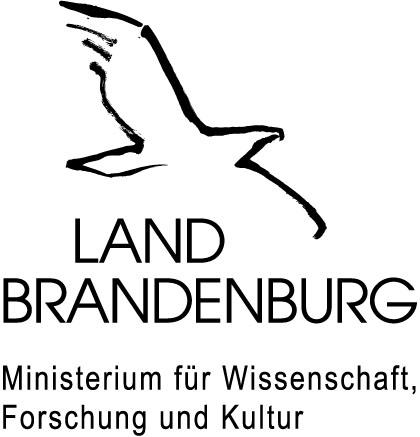}
\end{center}

\end{acknowledgement}

\providecommand{\latin}[1]{#1}
\makeatletter
\providecommand{\doi}
  {\begingroup\let\do\@makeother\dospecials
  \catcode`\{=1 \catcode`\}=2 \doi@aux}
\providecommand{\doi@aux}[1]{\endgroup\texttt{#1}}
\makeatother
\providecommand*\mcitethebibliography{\thebibliography}
\csname @ifundefined\endcsname{endmcitethebibliography}
  {\let\endmcitethebibliography\endthebibliography}{}

\end{document}